# New analytical forms of wave function in coordinate space and tensor polarization of deuteron

***V. I. Zhaba***

*Uzhgorod National University, Department of Theoretical Physics,*
*54, Voloshyna St., Uzhgorod, UA-88000, Ukraine*
*E-mail: viktorzh@meta.ua*

Numerically coefficients of the new analytical forms for deuteron wave function in coordinate space for NijmI, NijmII, Nijm93, Reid93 and Argonne v18 potentials are designed. The obtained wave functions do not contain superfluous knots. The designed parameters of a deuteron well agree with the experimental and theoretical data. The tensor polarization $t_{20}$ designed on wave functions is proportionate to earlier published outcomes.

## 1. INTRODUCTION

The deuteron is most the elementary nucleus, that will consist from two strongly interreacting particles (a proton and a neutron). Simplicity of a structure of a deuteron makes by its convenient laboratory for the study of nucleon-nucleon forces. Currently the deuteron is well investigated observationally and theoretically.

The experimental defined values of static performances of deuteron well agree with theoretical calculations. Despite of it, there are some theoretical inconsistencies. For example, one (for CD-Bonn potential) or both (for Moscow potential) components of the deuteron wave function have knots [1,2] about an origin of coordinates. Presence of the knots in wave functions of the basic and unique state of a deuteron testifies to inconsistencies and inaccuracies in embodying numerical algorithms in a solution of detailed problems. Influence of a select of numerical algorithms on solutions is reduced in Ref. [3,4].

Such potentials a nucleon-nucleon interaction as Bonn [1], Moscow [2], potentials of Nijmegen group (NijmI, NijmII, Nijm93 [5]), Argonne v18 [6] or Paris [7] potential have uneasy enough structure and bulky record. The original potential Reid68 was parameterized on the basis of phase analysis in Nijmegen group and has received title Reid93. The parametrization has been carried out for 50 parameters potential, and $\chi^2/N_{data}$=1.03 [5].

Besides deuteron wave function (DWF) can be submitted as the table: through corresponding files of values of radial wave functions. Sometimes at numerical calculations to operate with such files of numbers difficultly enough. And the text of programs for numerical calculations is overloaded. Therefore, it is feasible to obtain more simple analytical forms of representation DWF. It is possible to calculate the form factors and tensor polarization, characterizing the structure of the deuteron.

## 2. ANALYTICAL FORM OF THE DEUTERON WAVE FUNCTIONS

Known numerical values of a radial DWF in coordinate space can be approximated with the help of convenient expansions [8] in the analytical shape ($N_l$=11):

$$\begin{cases} u_1(r) = \sum_{i=1}^{N_1} A_i \exp(-a_i r^2), \\ w_1(r) = r^2 \sum_{i=1}^{N_1} B_i \exp(-b_i r^2). \end{cases}$$

To solve the associated system of Schrödinger equations that describe the radial DWF, back in 1955 y. it was proposed parameterization [9]:

$$\begin{cases} u = are^{-\mu r}, \\ w = bre^{-\mu r}, \end{cases} \qquad \begin{cases} u = ar^2 e^{-\mu r}, \\ w = br^3 e^{-\mu r}. \end{cases}$$

They can be generalized for the approximation of the DWF in the form of such analytical forms:

$$\begin{cases} u_2(r) = r^A \sum_{i=1}^{N_2} A_i \exp(-a_i r^3), \\ w_2(r) = r^B \sum_{i=1}^{N_2} B_i \exp(-b_i r^3). \end{cases} \tag{1}$$

At $N_2$=11 searching an index of function of a degree $r^n$ was carried out, that figures as a factor before the totals of exponential terms of the analytical shape (1). Best values appeared $n$=1.47 and $n$=1.01 for $u(r)$ and $w(r)$ accordingly. That is factors before the totals in (1) it is possible to pick as $r^{3/2}$ and $r^1$:

$$\begin{cases} u_2(r) = r^{3/2} \sum_{i=1}^{N_2} A_i \exp(-a_i r^3), \\ w_2(r) = r \sum_{i=1}^{N_2} B_i \exp(-b_i r^3). \end{cases} \tag{2}$$

Despite of unwieldy both long-lived calculations and minimizations $\chi^2$ (to size smaller for $10^{-4}$), it was necessary to approximate numerical values of wave functions of a deuteron, which arrays of numbers made 839x4 values to an interval $r$=0-25 fm for potential NijmI, NijmII, Nijm93 and Reid93 [5] and arrays of numbers made 1500x2 values to an interval $r$=0-15 fm for potential Argonne v18 [6].

The accuracy of the parametrization (2) is characterized:

$$\chi^2 = \frac{1}{n-p} \sum_{i=1}^{n} \left( y_i - f(x_i; a_1, a_2, ..., a_p) \right)^2, \tag{3}$$

where $n$ - the number of points $y_i$ of the array of the numerical values of DWF in the coordinate space; f - approximating function of $u$ (or $w$) according to formulas (2); $a_1, a_2, \ldots, a_p$ - parameters; $p$ - the number of parameters (coefficients in the sums of formulae (2)). Now, $\chi^2$ is determined not only by the shape of the approximating function $f$, but also the number of selections.

On known DWF (2) it is possible to calculate the deuteron properties (Table 1): the root-mean-square or matter radius $r_m$, the quadrupole moment $Q_d$, the magnetic moment $\mu_d$, the $D$- state probability $P_D$, the "$D/S$- state ratio" $\eta$. They well agree with the theoretical [6] and experimental [10] datas.

The designed DWF (2) do not contain superfluous knots (Fig.1). They well correlate with the data in Ref. [5]. The value of coefficients $A_i$, $a_i$, $B_i$, $b_i$ for formulas (2) is reduced in Tables 2-6.

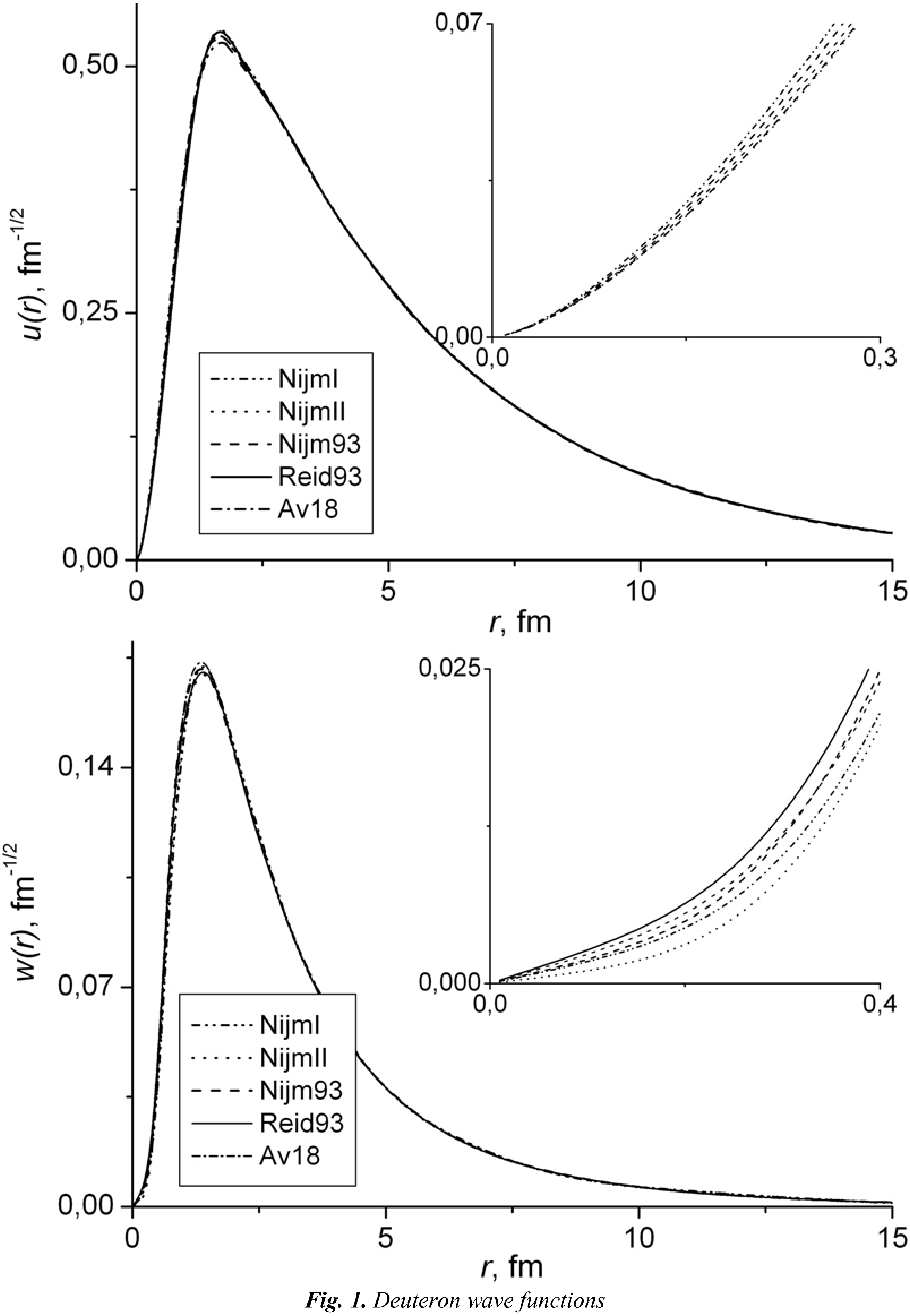

***Fig. 1.*** *Deuteron wave functions*

***Table 1.*** *Deuteron properties*

| | $P_D$ (%) | $r_m$ (fm) | $Q_d$ (fm$^2$) | $\mu_d$ ($\mu_N$) | $\eta$ |
|---|---|---|---|---|---|
| Nijm I (2) | 5.66274 | 1.96599 | 0.270883 | 0.847539 | 0.0285436 |
| Nijm I [5] | 5.664 | 1.967 | 0.2719 | - | 0.0253 |
| Nijm II (2) | 5.63002 | 1.96711 | 0.269793 | 0.847726 | 0.0277491 |
| Nijm II [5] | 5.635 | 1.968 | 0.2707 | - | 0.0252 |
| Nijm 93 (2) | 5.74956 | 1.96543 | 0.270572 | 0.847045 | 0.025257 |
| Nijm 93 [5] | 5.755 | 1.966 | 0.2706 | - | 0.0252 |
| Reid93 (2) | 5.69005 | 1.96811 | 0.270243 | 0.847384 | 0.024905 |
| Reid93 [5] | 5.699 | 1.969 | 0.2703 | 0.8853 | 0.0251 |
| Argonne v18 (2) | 5.75945 | 1.95401 | 0.268113 | 0.846988 | 0.025161 |
| Argonne v18 [6] | 5.76 | 1.967 | 0.270 | 0.847 | 0.0250 |
| Експеримент [10] | - | 1.975(3) | 0.2859(3) | 0.857438 | 0.0256(4) |

***Table 2.*** *Coefficients $A_i$, $a_i$, $B_i$, $b_i$* (NijmI)

| $i$ | $A_i$ | $a_i$ | $B_i$ | $b_i$ |
|---|---|---|---|---|
| 1 | 0.059474792604983 | 0.820048123485302 | 0.216703934160404 | -0.46062130918550 |
| 2 | 0.029815289846499 | 0.008925232016521 | 1.605852802910391 | -0.05815455970908 |
| 3 | 0.054093376131691 | 0.025498742322776 | 0.291264938712191 | 0.168130012367292 |
| 4 | 0.003754884398435 | 0.000853238833103 | 0.094286720550631 | 0.065723334248829 |
| 5 | 0.076152850036076 | 0.276725768486299 | 5.432373080615632 | -0.05688725511771 |
| 6 | 0.012623230198183 | 0.002868854515628 | 0.005309112421684 | 0.009307445117482 |
| 7 | 0.055227006777511 | 0.109973467568329 | 0.207373176910015 | 0.186164838918455 |
| 8 | 0.055367234187456 | 0.138446998423688 | 0.207372849758410 | 0.186596774580701 |
| 9 | 0.108440685265484 | 0.382785761360802 | 5.432373080319048 | -0.05688725495011 |
| 10 | 0.000542494001356 | 0.000217880096959 | 0.000831816055069 | 0.001352073835287 |
| 11 | 0.061459447768940 | 0.059212785941109 | 0.024128420330076 | 0.032026505193536 |

***Table 3.*** *Coefficients $A_i$, $a_i$, $B_i$, $b_i$* (NijmII)

| $i$ | $A_i$ | $a_i$ | $B_i$ | $b_i$ |
|---|---|---|---|---|
| 1 | 0.100534520422043 | 0.251018819916337 | 0.126943090633750 | -0.53113374184205 |
| 2 | 0.054286869011138 | 0.057442534595369 | 0.218940646725675 | 0.156510973923901 |
| 3 | 0.051278450057958 | 0.015600438117511 | 0.110684829492098 | 0.127632990059071 |
| 4 | 0.000814767311165 | 0.000260405824367 | 0.115497416682183 | 0.127617693287994 |
| 5 | 0.076562509307441 | 0.251007390539791 | 4.722499098968948 | -0.08308339515975 |
| 6 | 0.041994669536818 | 0.251009303096756 | 0.004871680145179 | 0.008411658752901 |
| 7 | 0.005869159261571 | 0.001146927630628 | 0.117531418761814 | 0.127621420370460 |
| 8 | 0.033972775878515 | 0.057451551324434 | 0.116012967954140 | 0.127618151426690 |
| 9 | 0.048710359630614 | 0.251011494582336 | 4.722499098969801 | -0.08308339516002 |
| 10 | 0.019984994410513 | 0.004345797074014 | 0.000776726189794 | 0.001208526589511 |
| 11 | 0.031082972309765 | 0.057443362285818 | 0.022141456306319 | 0.030108870003356 |

***Table 4.*** *Coefficients $A_i$, $a_i$, $B_i$, $b_i$* (Nijm93)

| $i$ | $A_i$ | $a_i$ | $B_i$ | $b_i$ |
|---|---|---|---|---|
| 1 | 0.108158922056900 | 0.322843406882508 | 0.179007868039459 | -0.49551653602562 |
| 2 | 0.059853308401022 | 0.322238702916314 | 0.254707539462443 | 0.224767666483098 |
| 3 | 0.000988499937104 | 0.000284873387262 | 0.012705460633378 | 0.014729943005680 |
| 4 | 0.048517972009155 | 0.075231693391126 | 0.001144517190176 | 0.001283074302584 |
| 5 | 0.062564722993285 | 0.322189096838678 | 5.013179412294758 | -0.08329753146954 |
| 6 | 0.007194322349655 | 0.001326489490838 | 0.154332010904778 | 0.199804896403292 |
| 7 | 0.062095388454927 | 0.020017090383419 | 0.000262490061461 | 0.000166238317782 |
| 8 | 0.049196871063984 | 0.073208280814316 | 0.004039748088559 | 0.005059646459065 |
| 9 | 0.024533675856846 | 0.005315783506284 | 5.013179410725441 | -0.08329753146557 |
| 10 | 0.038064498676696 | 0.076574723953854 | 0.154332010904783 | 0.199804896403214 |
| 11 | 0.036149451152707 | 0.322417369499423 | 0.038551683049846 | 0.034958655682995 |

***Table 5.*** *Coefficients $A_i$, $a_i$, $B_i$, $b_i$* (Reid93)

| $i$ | $A_i$ | $a_i$ | $B_i$ | $b_i$ |
|---|---|---|---|---|
| 1 | 0.069165902087054 | 0.260694028992927 | 0.046691520051938 | -0.45747046709213 |
| 2 | 0.016816805243013 | 0.003745637363879 | 0.251783375231406 | 0.090050028918338 |
| 3 | 0.036276406025602 | 0.024248992643587 | 0.009532804413578 | 0.011344840155198 |
| 4 | 0.054276523719886 | 0.260700699439171 | 0.000923945728241 | 0.000912747409340 |
| 5 | 0.090259458024542 | 0.260967752832736 | 5.271594771417672 | -0.07565077617374 |
| 6 | 0.061396673091335 | 0.063889193541223 | 0.052806780918954 | 0.183289904460344 |
| 7 | 0.031841769081845 | 0.011654989924025 | 0.000228716768808 | 0.000121831213912 |
| 8 | 0.005059996000766 | 0.001034701614202 | 0.003090018155478 | 0.003637935243084 |
| 9 | 0.049562672796619 | 0.260727250929229 | 5.234984210722409 | -0.07564966175244 |
| 10 | 0.000707435476194 | 0.000244188362049 | 0.052806780918950 | 0.183289904460260 |
| 11 | 0.052754370232542 | 0.063865550323201 | 0.037043195071994 | 0.161860429844891 |

***Table 6.*** *Coefficients $A_i$, $a_i$, $B_i$, $b_i$* (Argonne v18)

| $i$ | $A_i$ | $a_i$ | $B_i$ | $b_i$ |
|---|---|---|---|---|
| 1 | -2.131341252505580 | 0.031643712686560 | 0.080647832829630 | 1.093497500081610 |
| 2 | 0.002862340193692 | 0.000542907221112 | 4.276423111508590 | -0.164408062648391 |
| 3 | 0.256334694145763 | 0.294291730249029 | 0.000624941838599 | 0.000754447617320 |
| 4 | 0.423402972585148 | 0.031312338359191 | 0.009727269524289 | 0.011969072757668 |
| 5 | 0.145430732510713 | 0.069905808009759 | 0.301941164444362 | 0.042806729356171 |
| 6 | 0.011414515917362 | 0.002346721872793 | 0.080072843315215 | -0.426032563900697 |
| 7 | 0.462503468645901 | 0.031163927385390 | 0.030479793609307 | 0.030097619538244 |
| 8 | 0.429637798877398 | 0.031303117915512 | 0.002855363369814 | 0.003778414883522 |
| 9 | 0.403306170391361 | 0.031326954270918 | 0.301739482874308 | 0.042806599897975 |
| 10 | 0.029251099139079 | 0.007877840866271 | 0.080072843315210 | -0.426032563900809 |
| 11 | 0.447281105021924 | 0.031253678455058 | 0.080082395462851 | -0.186678441085246 |

## 3. ФORMFAKTORS AND TENZOR POLARIZATION OF THE DEUTERON

Measuring of polarization characteristics of a response of a fragmentation of deuteron *A(d,p)X* at the intermediate and high energies remains to one of the basic tools for examination of structure of a deuteron. For a quantitative understanding of the structure of the deuteron *S* - and *D* - States and polarization characteristics are considered different models of the nucleon-nucleon potential. The charge distribution of the deuteron is not well known from experiment, because it is only through the use of polarization measurements, and the scattered unpolarized elastic differential cross sections [11-14]. However, it can be defined [11]. Differential cross section of elastic scattering of unpolarized electrons by unpolarized deuterons without measuring the polarization of the repulsed electrons and deuterons [12,13]

$$\frac{d\sigma}{d\Omega}=\left(\frac{d\sigma}{d\Omega}\right)_{MOTT}S,$$

$$S=A(p)+B(p)\tan^2\left(\frac{\theta}{2}\right).$$

Here $\theta$ - the scattering angle in the laboratory system, $p$ - the momentum of the deuteron in the fm$^{-1}$, *A(p)* and *B(p)* - function of the electric magnetic structure [12,13]:

$$A(p)=F_C^2(p)+\frac{8}{9}\eta^2F_Q^2(p)+\frac{2}{3}\eta F_M^2(p),\tag{4}$$

$$B(p)=\frac{4}{3}\eta(1+\eta)F_M^2(p),\tag{5}$$

where $\eta=\frac{p^2}{4M_D^2}$; $M_D$=1875.63 MeV - deuteron mass. Charge $F_C(p)$, quadrupole $F_Q(p)$ and magnetic $F_M(p)$ form factors contain information about the electromagnetic properties of the deuteron [11-14]:

$$F_C=\left[G_{Ep}+G_{En}\right]\int_0^\infty\left[u^2+w^2\right]j_0dr;\tag{6}$$

$$F_Q=\frac{2}{\eta}\sqrt{\frac{9}{8}}\left[G_{Ep}+G_{En}\right]\int_0^\infty\left[uw-\frac{w^2}{\sqrt{8}}\right]j_2dr;\tag{7}$$

$$F_M=2\left[G_{Mp}+G_{Mn}\right]\int_0^\infty\left[\left(u^2-\frac{w^2}{2}\right)j_0+\right.$$
$$\left.+\left(\frac{uw}{\sqrt{2}}+\frac{w^2}{2}\right)j_2\right]dr+\frac{3}{2}\left[G_{Ep}+G_{En}\right]\int_0^\infty w^2\left[j_0+j_2\right]dr;\tag{8}$$

where $u$, $w$ - radial DWF (2), $j_0$, $j_2$ - the spherical Bessel function of the argument $pr/2$; $G_{Ep}$, $G_{En}$ ($G_{Mp}$, $G_{Mn}$) - neutron and proton electric (magnetic) form factors. In the experiments with unpolarized elastic scattering the structure functions can be obtained by determining *B(p)* directly from the cross-sectional dispersion ago. Tensor polarization of deuterons repulsed determined [11-13] using the form factors (6)-(8):

$$t_{20}(p)=-\frac{1}{\sqrt{2}S}\left(\frac{8}{3}\eta F_C(p)F_Q(p)+\frac{8}{9}\eta^2F_Q^2(p)+\frac{1}{3}\eta\left[1+2(1+\eta)tg^2\left(\frac{\theta}{2}\right)\right]F_M^2(p)\right),\tag{9}$$

$$t_{21}(p)=\frac{2}{\sqrt{3}S\cos\left(\frac{\theta}{2}\right)}\eta\sqrt{\eta+\eta^2\sin^2\left(\frac{\theta}{2}\right)}F_M(p)F_Q(p),\tag{10}$$

$$t_{22}(p) = -\frac{1}{2\sqrt{3}S}\eta F_M^2(p). \quad (11)$$

The calculation of the polarization tensor (9) and (10) for the NijmI, NijmII, Nijm93, Reid93 and Argonne v18 potentials (Fig. 2-4) and compared the obtained results with the literature experimental and theoretical data.

The effect of the accuracy of the $\chi^2$ approximation (2) for DWF to calculate the tensor polarization $t_{20}(p)$ at scattering angle $\theta$=70$^0$ for Argonne v18 potential (Fig. 2). Compared $t_{20}(p)$, the two obtained approximations. The value of $\chi^2$ of these approximations for $u(r)$ and $w(r)$ are $10^{-6}$ and $10^{-6}$, $10^{-9}$ and $10^{-10}$, respectively. It turned out that the "worst" approximation significantly affects the result of the calculation is $t_{20}(p)$. For example, when the momentum of 3.6 fm$^{-1}$ between approximations *1* and *2* is 2.8%.

A detailed comparison of the obtained values of $t_{20}(p)$ (the scattering angle θ=70$^0$) for NijmI, NijmII, Nijm93, Reid93 and Argonne v18 potentials (Fig. 3) with modern experimental data of JLAB t20 [11,14] and BLAST [15,16] collaboration. Good agreement for the momentum $p$=1-4 fm $^{-1}$. The calculated value of $t_{20}(p)$is well agreed with the results of the work, where the theoretical calculations: data [15] for the Paris, Argonne v14 and Bonn-E potentials and with the data of [17] to Moscow, NijmI, NijmII, CD-Bonn, and Paris potentials, as well as with the values [10] for elastic *ed*-scattering models with the inclusion of nucleon isobaric components, within light-front dynamics and quark cluster model, for the Bonn-A, Bonn-B and Bonn-C, Bonn Q, Reid-SC, and Paris A-VIS potential. Good agreement between the obtained values of $t_{21}(p)$ data for models and NRIA and NRIA+MEC+RC [15]. Besides $t_{20}(p)$ and $t_{21}(p)$coincide well with the results according to effective field theory [16].

In the scientific literature missing experimental data for $t_{21}(p)$ and $t_{22}(p)$ in a wide range of momentums. So is the actual receipt of these variables both theoretically and experimentally. Are also appropriate calculations of the polarization characteristics of the deuteron (tensor components of the sensitivity to the polarization of the deuteron $T_{20}$, polarization transmission $K_0$ and the tensor analyzing power $A_{yy}$) and their comparison with theoretical calculations [4] and also with experimental data [18].

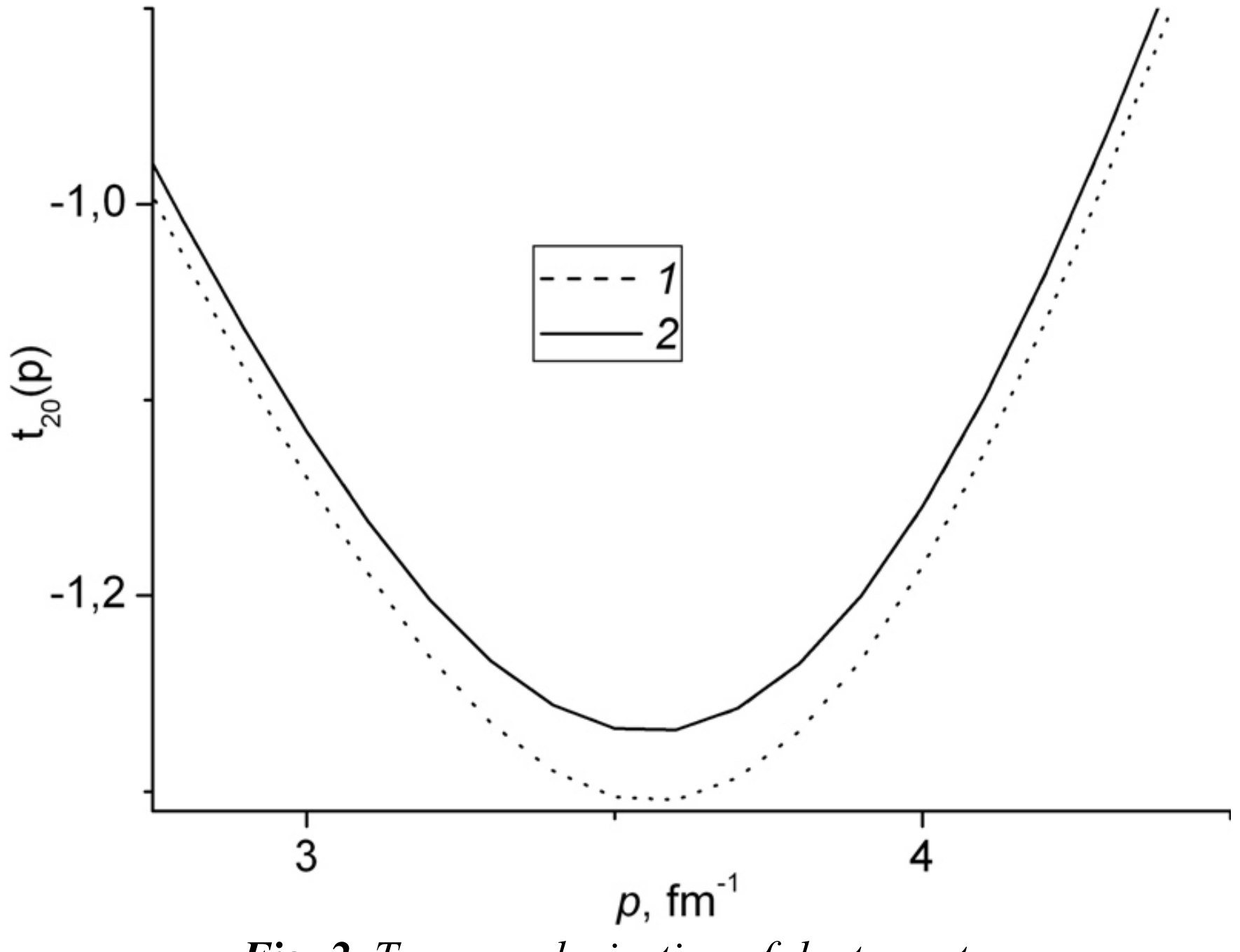

***Fig. 2.*** *Tensor polarization of deuteron $t_{20}$*

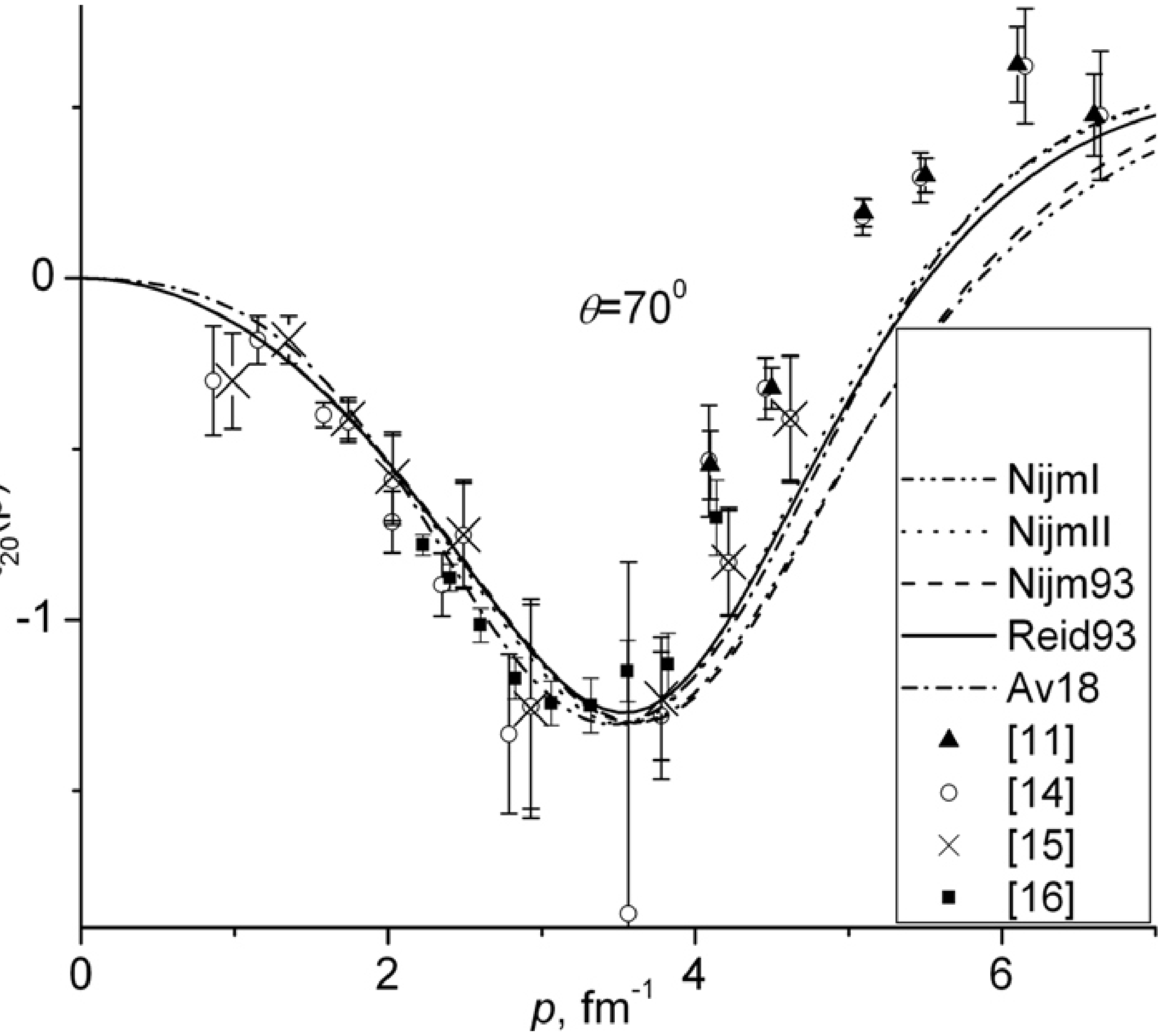

***Fig. 3.*** *Tensor polarization of deuteron* $t_{20}$

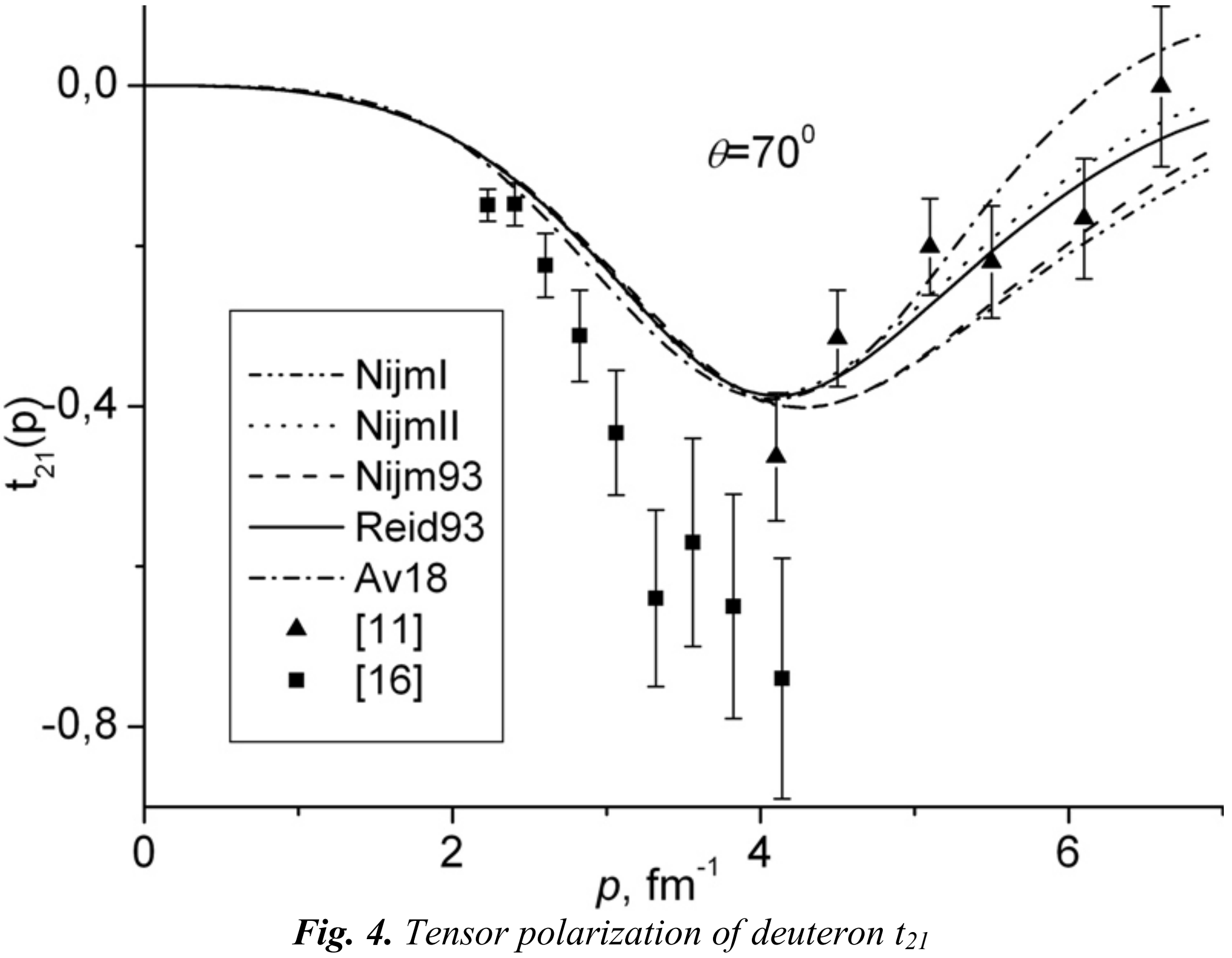

***Fig. 4.*** *Tensor polarization of deuteron* $t_{21}$

## CONCLUSIONS

The coefficients of the approximating dependences in an new analytic form (2) for the numerical values of DWF in the coordinate space for realistic phenomenological NijmI, NijmII, Nijm93, Reid93 and Argonne v18 potentials. The behavior of $\chi^2$ value depending on the number of summands of the decomposition of $N_i$. Taking into account the minimum values of $\chi^2$ for these forms built DWF in the coordinate representation, which does not contain redundant knots. The calculated parameters of the deuteron are well agreed with theoretical and experimental results.

On received DWF calculated deuteron tensor polarization. Numerical calculations of the deuteron tensor polarization $t_{20}(p)$ and $t_{21}(p)$ carried out in the range of momentum 0-7 $fm^{-1}$. The result $t_{20}(p)$ for NijmI, NijmII, Nijm93, Reid93 and Argonne v18 potentials agreed well with the literature results for other potential nucleon-nucleon models, and with experimental data.

The results obtained deuteron tensor polarization $t_{ij}(p)$ give some information about the electromagnetic structure of the deuteron differential cross section and double scattering, if there was known of the tensor analyzing power.

## References

1. R. Machleidt. High-precision, charge-dependent Bonn nucleon-nucleon potential // *Phys. Rev. C*. 2001, v.63, p.024001-024032.
2. V.I. Kukulin, V.N. Pomerantsev, A. Faessler et al. Moscow-type NN-potentials and three-nucleon bound states // *Phys. Rev. C*. 1998, v.57, p.535-554.
3. I. Haysak, V. Zhaba. On the nods of the deuteron wave function // *Visnyk Lviv Univ. Ser. Phys.* 2009, N.44, C. 8-15.(in Ukrainian).
4. I.I. Haysak, V.I. Zhaba. Deuteron: wave function and parameters // *Uzhhorod Univ. Scien. Herald. Ser. Phys.* 2014, N. 36, C. 100-106.(in Ukrainian).
5. V.G.J. Stoks, R.A.M. Klomp, C.P.F. Terheggen et al. Construction of high quality NN potential models // P*hys. Rev. C*. 1994, v.49, p.2950-2962; J.J. de Swart, R.A.M.M. Klomp, M.C.M. Rentmeester et al. The Nijmegen Potentials // *Few-Body Systems*. 1995, v. 8, p. 438-447.
6. R.B. Wiringa, V.G.J. Stoks, R. Schiavilla. Accurate nucleon-nucleon potential with charge-independence breaking // *Phys. Rev. C*. 1995, v.51, p.38-51.
7. M. Lacombe, B. Loiseau, R. Vinh Mau et al. Parametrization of the deuteron wave function of the Paris N-N potential // *Phys. Lett. B*. 1981, v.101, p.139-140.
8. S.B. Dubovichenko. *Properties of light atomic nuclei in the potential cluster model*. Almaty: “Daneker”, 2004, 247 p.(in Russian).
9. F. Cap, W. Gröbner. New method for the solution of the deuteron problem, and its application to a regular potential // *Nuovo Cimento*. 1955, v.1, p. 1211-1222.
10. M. Garcon, J.W. van Orden. The deuteron: structure and form factors // *Adv. Nucl. Phy*s. 2001, v. 26, p. 293-378.
11. D. Abbott et al. Measurement of Tensor Polarization in Elastic Electron-Deuteron Scattering at Large Momentum Transfer // *Phys. Rev. Lett*. 2000, v. 84, p.5053-5057.
12. R. Gilman, F. Gross. Electromagnetic structure of the deuteron // *J. Phys. G*. 2002, v. 28, p. R37-R116.
13. A.K.A. Azzam, M.A. Fawzy, E.M. Hassan et al. Electron-Deuteron Tensor Polarization and D-State Probability // *Turk. J. Phys*. 2005, v. 29, p. 127-135.
14. D. Abbott et al. Phenomenology of the deuteron electromagnetic form factors // *Eur. Phys. J. A*. 2000, v. 7, p. 421-427.
15. M. Garson, J. Arvieux, D.H. Beck et al. Tensor polarization in elastic electron-deuteron scattering in the momentum transfer range $3.8 \leq Q \leq 4.6$ $fm^{-1}$ // *Phys. Rev. C*. 1994, v. 49, p. 2516.
16. C. Zhang, M. Kohl, T. Akdogan et al. Precise Measurement of Deuteron Tensor Analyzing Powers with BLAST // *Phys. Rev. Lett*. 2011, v. 107, p. 252501.
17. N.A. Khokhlov, A.A. Vakulyuk. Elastic Electron-Deuteron Scattering within a Relativistic Potential Model // *Phys. Atom. Nucl.* 2015, v.78, p.92-104.
18. L.S. Azhgirey, S.V. Afanasiev, A.Yu. Isupov et. al. Measurement of the tensor Ayy and vector Ay analyzing powers of the deuteron inelastic scattering of beryllium at 5.0 GeV/c and 178 mrad // *Yad. Fiz.* 2005, v.68, p.1029-1036.(in Russian).